\documentclass[11pt,a4paper]{article}

\usepackage{graphicx}
\usepackage{latexsym}
\usepackage{amsmath,amsfonts,amsthm}

\renewcommand{\baselinestretch}{1.3} 

\renewcommand{\theequation}{\thesection.\arabic{equation}}

\begin{document}
\pagenumbering{roman} \setcounter{page}{0} 
\vspace{-2cm}
\title{Monte Carlo Investigation of Lattice Models of Polymer Collapse in Five
  Dimensions}
\author{ A. L. Owczarek\ddag \, and T. Prellberg\dag  \thanks{{\tt {\rm email:}
aleks@ms.unimelb.edu.au,thomas.prellberg@tu-clausthal.de}} \\
        \ddag Department of Mathematics and Statistics,\\
         The University of Melbourne, 3010, Australia.\\
\dag Institut f\"ur Theoretische Physik,\\
        Technische Universit\"at Clausthal,\\ Arnold Sommerfeld Stra\ss e 6,\\
        D-38678 Clausthal-Zellerfeld, Germany
 }
\date{\today
}
\maketitle 
 
\begin{abstract} 
  
  Monte Carlo simulations, using the PERM algorithm, of interacting
  self-avoiding walks (ISAW) and interacting self-avoiding trails (ISAT) in
  five dimensions are presented which locate the collapse phase transition in
  those models. It is argued that the appearance of a transition (at least) as
  strong as a \emph{pseudo-first-order} transition occurs in both models.
  The values of various theoretically conjectured dimension-dependent
  exponents are shown to be consistent with the data obtained.  Indeed the
  first-order nature of the transition is even stronger in five dimensions
  than four.  The agreement with the theory is better for ISAW than ISAT and
  it cannot be ruled out that ISAT have a true first-order transition in
  dimension five. This latter difference would be intriguing if true. On the
  other hand, since simulations are more difficult for ISAT than ISAW at this
  transition in high dimensions, any discrepancy may well be due to the
  inability of the simulations to reach the true asymptotic regime.

\vspace{1cm} 

\noindent{\bf Short Title:} 5d lattice polymer collapse


\end{abstract} 

\vfill

\newpage

\pagenumbering{arabic}
\section{Introduction} 
\label{intro}
\setcounter{page}{1}

Recently, evidence \cite{prellberg2000a-:a,owczarek2000a-:a,prellberg2001a-:a}
has been presented, from the investigation of various four-dimensional lattice
models, that the collapse, or coil-globule, transition of an isolated polymer
in solution can be seen in high-dimensional models. Moreover, while being
second-order, the transition elucidated seems to display many of the
characteristics of a first-order transition. At first sight the importance of
these findings for physical polymers may not be apparent, but one must
remember that in lower dimensions it has been well established that polymer
collapse can be described by a tricritical O(0) field theory with an upper
critical dimension of three
\cite{gennes1975a-a,stephen1975a-a,duplantier1982a-a}. So extending the
evidence for these findings in high dimensions will eventually require a new
understanding of the finite size scaling associated with the tricritical
theory.

As a consequence of the above considerations, a theoretical framework has been
conjectured by the present authors \cite{prellberg2000a-:a,owczarek2000a-:a}.
This framework keeps some aspects of the expected behaviour in the infinite 
polymer limit and is the most likely to fit all the evidence available at present. 
The needed framework has been provided by re-evaluating the validity and meaning 
of the older collapse transition theory of Lifshitz, Grosberg and Khokhlov
\cite{lifshitz1968a-a,lifshitz1976a-a,lifshitz1978a-a,khokhlov1981a-a} in the
light of modern developments and restricting this older theory to dimensions
above the upper critical dimension.  The phase transition, which should still
be a second-order transition in the thermodynamic limit, now has
finite-polymer-length scaling behaviour with first-order characteristics, and
so has been named a \emph{pseudo-first-order} (PFO) transition. The proposed
theory also predicts various dimension-dependent exponents while the general
scenario of `false' first-order behaviour should be seen in any dimension
greater than three.  Drawing upon Monte Carlo simulations using the PERM
algorithm, we confirm here that indeed the general scenario occurs in five
dimensions as well as four. Both interacting self-avoiding walks (ISAW) and
interacting self-avoiding trails (ISAT) on a five-dimensional hyper-cubic
lattice have been simulated. Clear bimodal distributions for the internal
energy are found, that become more distinct with increasing polymer length.
Indeed the first-order nature of the transition is even stronger in five
dimensions than four. The agreement with the PFO theory is better for ISAW
than ISAT, and we cannot rule out that ISAT (or even completely for ISAW) have
a true first-order transition in the thermodynamic limit in dimension five.
This latter difference would be intriguing if true though we are inclined to
argue that since simulations are more difficult for ISAT than ISAW at this
transition in high dimensions, any discrepancy is due to the inability of the
simulations to reach the true asymptotic regime.  Regardless, our results do
imply that the transition is at least as strong as the PFO theory predicts and
a dimensional dependence occurs for the associated exponents. If a true
thermodynamic first-order transition does occur then the tricritical theory
mentioned above would need substantial revision.
  
We begin by summarising the predictions of the PFO transition theory and by reviewing 
the previous work in four dimensions in the next section. In Section 3 we present
the results of our simulation of five-dimensional ISAW and ISAT.

\section{Review}
\label{review}

The collapse transition describes the phase transition of an isolated polymer
in dilute solution from a high temperature state, which in low dimensions is
dominated by excluded volume effects, to a low temperature state dominated by the
attractive interactions between monomers so that the polymer forms a dense
globule like a liquid drop. The temperature of the phase
transition has been known as the $\theta$-point and signifies a change in the
scaling of the mean-square radius of gyration with polymer length $N$,
\begin{equation}
  R^2_{g,N} \sim a \: N^{2\nu}\quad\mbox{as}\quad N\rightarrow\infty\; ,
\end{equation}
such that for low temperatures $\nu = 1/d$ in dimension $d$. In three
dimensions, for example, $\nu \approx 0.5874(2)$ at high temperatures
\cite{prellberg2001b-a}, $\nu =1/2$ (the random walk value) at the
$\theta$-point, and $\nu = 1/3$ at low temperatures. However, the upper
critical dimension of the excluded volume state is expected to be four so that
in five dimensions any excluded volume state will behave, to first
approximation, in the same fashion as a pure random walk and in the same
fashion as the $\theta$-point, which is expected to have an upper critical
dimension of three, with $\nu =1/2$. So in high dimensions the nature of the
collapse apparently changes somewhat with a gross change between two states: a
random walk or Gaussian state at high temperatures and a low temperature
globular state with $\nu=1/5$ in five dimensions.

On a deeper level, as recently shown \cite{owczarek2001a-:a} the excluded
volume effects at high temperatures do not disappear altogether, and reappear
as corrections-to-scaling. There is then a subtle sub-dominant difference
between the excluded volume state and $\theta$-state (which after-all is defined as
the point where the excluded volume effects are cancelled out by the
attractive forces between monomers). Hence, this provides a method for locating
the $\theta$-point.

If we now shift our attention to the thermodynamic limit,
the application of standard mean-field theory would predict a second-order
phase transition with a jump in the specific heat. For finite polymer length
one may naively  expect that a crossover occurs in a range of temperatures of
the order of $N^{-1/2}$, that is the crossover exponent is $1/2$ regardless of
dimension ($d\geq 3$).

Now, the theory \cite{khokhlov1981a-a,prellberg2000a-:a,owczarek2000a-:a} of
the \emph{pseudo-first-order} transition also predicts a thermodynamic
second-order transition at a Gaussian $\theta$-point with a jump in the
specific heat. However, for finite polymer length the situation differs from
the naive theory above. The size of the crossover (or rounding) region of the
transition is asymptotically small relative to the shift of the transition.
Inside the crossover region the transition also takes on the characteristics
of a first-order transition when considered at finite polymer length. That is,
if one considers the distribution of the internal energy at fixed polymer
length, then a double peaked distribution occurs in the transition region that
becomes sharper with increasing polymer length. In fact, a well defined latent
heat can be ascribed. However, this latent heat goes to zero in the
thermodynamic limit.

The consequences of the theory are that the polymer collapse transition in
high dimensions is shifted
below the $\theta$-point by a temperature of the order of
$O(N^{-1/(d-1)})$. Using an effective Boltzmann weight $\omega=e^{J/k_BT}$, where
$-J$ is the energy associated with a single nearest-neighbour interaction for ISAW, or 
a single contact for ISAT, respectively, a finite-size transition temperature\footnote{for the sake of ease of expression
  in this section we will use the word ``temperature'' to mean the effective
  Boltzmann weight} $\omega_{c,N}$ approaches the $\theta$-temperature
($\omega_\theta$) as
\begin{equation}
\label{shift}
\omega_{c,N} -\omega_\theta  \sim \frac{s}{N^{1/(d-1)}}
\end{equation}
for some constant $s$. That is, the polymer collapse \emph{shift}
exponent is $1/4$ in five dimensions.  The width of the transition region $\Delta \omega$
at finite $N$ is predicted  to scale as
\begin{equation}
\label{width}
\Delta \omega \sim \frac{w}{N^{(d-2)/(d-1)}}
\end{equation}
for some constant $w$. That is, the polymer collapse \emph{crossover} exponent
is $3/4$ in five dimensions. Hence as mentioned above the size of the
crossover region is asymptotically small relative to the shift of the
transition. Over the width of the transition there is a rapid change in the internal
energy that scales as $O(N^{-1/(d-1)})$: the important point here of course is
that this tends to zero for infinite length so the effect of the peak in the
specific heat is scaled away for $N$ large, leaving a finite jump in the
thermodynamic limit.  Just as important, as mentioned above, is to consider
the full distribution of internal energy $\rho_{\omega,N}(E)$ as a function of 
temperature $\omega$ and polymer length $N$. For any $\omega$ below $\omega_\theta$ (high temperatures),
and those well above $\omega_{c,N}$ (low temperatures),
one expects the distribution of internal energy to look like a single peaked
distribution centred close to the thermodynamic limit value: a Gaussian
distribution is expected around the peak with variance $O(N^{-1/2})$. In fact,
this picture should be valid for all temperatures outside the range $[\omega_{c,N}
- O(N^{-(d-2)/(d-1)}),\omega_{c,N} + O(N^{-(d-2)/(d-1)})]$. When this region is
entered one expects to see a double peaked distribution as in a first-order
transition region. For any temperature in this region there should be two
peaks in the internal energy distribution separated by a gap $\delta U$ of the
order of $O(N^{-1/(d-1)})$. Each peak should be of
Gaussian type with individual variances again of the order of $O(N^{-1/2})$. 
Defining the ``interfacial tension as the height of the minimum
between the two peaks relative to the height of the maxima (at an
appropriately chosen temperature, at which both peaks are of equal height),
this scaling implies an exponential decrease of the interfacial tension 
in $N(\Delta U)^2$, i.e. as $\exp(-\mbox{const}\,N^{(d-3)/(d-1)})$.
Hence as $N$ increases the peaks will become more and more distinct and
relatively sharper but the peak positions will be getting closer together.
Hence this scenario has been referred to as a \emph{pseudo}-first-order
transition. If there were a real first-order transition then the distance
between the peaks should converge to a non-zero constant. On the other hand
the transition is \emph{not} a conventional second-order phase transition with
a well defined limit distribution of the internal energy that is simply
bimodal.

In previous work this above scenario has been established for the lattice
models of self-avoiding walks \cite{prellberg2000a-:a,owczarek2000a-:a}
interacting via nearest neighbour attractive potentials on the
four-dimensional hyper-cubic lattice and self-avoiding trails (lattice paths
that are bond-avoiding but not site avoiding) \cite{prellberg2001a-:a}
interacting via site contact potentials on the same lattice. The first-order
nature was well established with a reasonable fit to the shift and crossover
exponents found. The latent heat was, if anything, larger than expected though
it was shown to be decreasing, as expected, with increasing polymer length
(two different measures of the latent heat were compared for the sake of
consistency).

In this work we have considered both self-avoiding walks and trails
interacting as described above on five-dimensional hyper-cubic lattices.
This allows us to, firstly, confirm that the above results were not peculiar
to four dimensions, and, secondly, to attempt to confirm the dimensional
dependence of the shift and crossover of the transition, namely that the shift
becomes larger and the crossover sharper as the dimension is increased.
Expecting these results however means that while the transition will be easier
to see, it will be more difficult to simulate because of the inherent difficulty
in computer simulation of any first-order like transition.

\section{Results}
\label{results}

We have simulated ISAW and ISAT on a five-dimensional hyper-cubic lattice using
the Pruned-Enriched Rosenbluth Method (PERM), a clever generalisation
of a simple kinetic growth algorithm
\cite{grassberger1997a-a,frauenkron1998a-a}.  PERM builds upon the
Rosenbluth-Rosenbluth method \cite{rosenbluth1955a-a}, in which walks/trails
are generated by simply growing an existing walk/trail kinetically, but
overcomes the exponential ``attrition'' and re-weighting needed in
this approach by a combination of enrichment and pruning strategies. 
Our implementation here directly extends our previous ISAW and ISAT work
\cite{prellberg2000a-:a,owczarek2000a-:a,prellberg2001a-:a}. 

Each run had a maximum length $N_{max}$ set and while individual runs
gave information about shorter lengths we collected data from
independent runs at some shorter lengths to guarantee statistical
independence.  For ISAW, simulations were conducted with the maximum lengths
$N_{max}$ set to 256, 384, 512, 768, 1024, 1536, and 2048, and for ISAT, simulations
were conducted with the maximum lengths $N_{max}$ set to 64, 96, 128, 192, 256, 384, and 512.
The values of $\omega$ were chosen to cover double the width of the collapse region. 
For this, we estimated the approximate position and width of the specific heat peak at each $N_{max}$
with an initial simulation and then ran extended simulations at five temperatures around the peak 
position covering this range.

To locate the $\theta$-point, we ran many closer spaced simulations for ISAW 
in the range of $\omega$ from $1.1$ to $1.2$ at length $N_{max}=16384$, and for 
ISAT in the range of $\omega$ from $1.26$ to $1.32$ at length $N_{max}=16384$.
At each fixed $\omega$, we generated $10^8$ configurations of maximal length.
To illustrate the computational effort, the generation of a
sample of size $10^8$ at length $N_{max}=16384$ took several months
CPU time on a $1.3$ GHz Pentium-3 Xeon. 

We computed statistics for $R_{e,N}^2$, the average square of the end-to-end distance, and for $R_{m,N}^2$, 
the average square of the mean distance of the internal sites of the walk from the end points 
for an $N$-step walk respectively trail, the partition function
$Z_N$, the internal energy $U_N$ and specific heat $C_N$. Moreover, we
generated the distribution of the number of interactions at $N_{max}$. The
distributions obtained at various temperatures were then combined using the
multiple histogram method \cite{ferrenberg1988a-a}. Error bars were computed
as previously described \cite{prellberg2000a-:a,prellberg2001a-:a}.

\begin{figure}
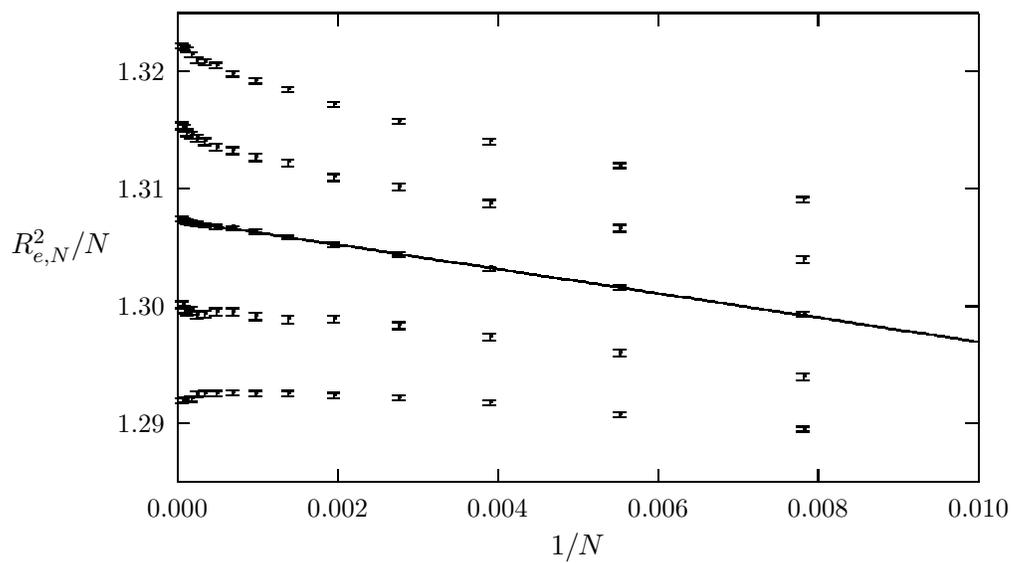

\begin{center}
% GNUPLOT: LaTeX picture
\setlength{\unitlength}{0.240900pt}
\ifx\plotpoint\undefined\newsavebox{\plotpoint}\fi
\sbox{\plotpoint}{\rule[-0.200pt]{0.400pt}{0.400pt}}%

\end{center}
\caption{\it $R_{e,N}^2/N$ versus $1/N$ in the $\theta$-region with
$\omega=1.120$, $1.125$, $1.130$, $1.135$, $1.140$ from top to bottom.
The linear fit for $\omega=1.130$ indicates a clear $1/N$-correction.}
\label{figure1} 
\end{figure}

\begin{figure}
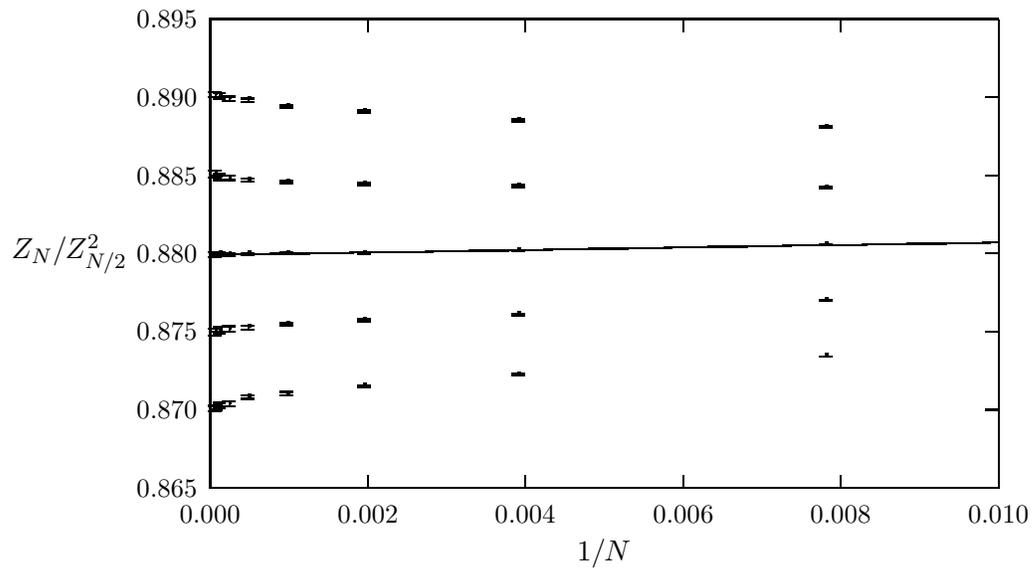

\begin{center}
% GNUPLOT: LaTeX picture
\setlength{\unitlength}{0.240900pt}
\ifx\plotpoint\undefined\newsavebox{\plotpoint}\fi
\sbox{\plotpoint}{\rule[-0.200pt]{0.400pt}{0.400pt}}%
%
\end{center}
\caption{\it $Z_N/Z_{N/2}^2$ versus $1/N$ in the $\theta$-region with
$\omega=1.120$, $1.125$, $1.130$, $1.135$, $1.140$ from bottom to top.
The linear fit for $\omega=1.130$ indicates a clear $1/N$-correction.}
\label{figure2} 
\end{figure}

As indicated above, recent evidence \cite{owczarek2001a-:a} has shown that the
corrections to scaling in the swollen phase in five dimensions lead to a
scaling of the size of the polymer as
\begin{equation}
R^2_N\sim dN\left(1+cN^{-1/2}+O(N^{-1})\right) 
\end{equation}
and to a scaling of the partition function as
\begin{equation}
Z_N\sim a\mu^N\left(1+bN^{-1/2}+O(N^{-1})\right)\;.
\label{ZN}
\end{equation}
We argued in \cite{owczarek2001a-:a} that the source of the $N^{-1/2}$
corrections is the excluded volume effect.  Therefore, these should disappear
at the $\theta$-point. Hence, one method to locate the $\theta$-point is to
find where such corrections vanish. Figure \ref{figure1} shows that the
$N^{-1/2}$ corrections vanish near $\omega=1.13$ for the end-to-end distance
of ISAW. Figure \ref{figure2} shows the scaling of $Z_N/Z_{N/2}^2$ for ISAW,
which, as a consequence of Equation (\ref{ZN}), scales as
$a^{-1}\left(1-b(\sqrt2-1)N^{-1/2}+O(N^{-1})\right)$. Again, we find that the
correction to scaling vanishes near $\omega=1.13$. The consistency of the two
estimates is a validation of our method.  From our data, we estimate the
location of the ISAW $\theta$-point to be $\omega_\theta=1.130(5)$. For ISAT,
we obtain analogously a $\theta$-point estimate of $\omega_\theta=1.29(2)$.

At this point it is intriguing to note that the PERM algorithm is most efficient at, or near, the $\theta$-point.
If one could indeed use this efficiency criterion to locate the $\theta$-point, this would, for example, lead to a very precise 
estimate of $1.1305(10)$ for the $\theta$-point of ISAW.

\begin{figure}
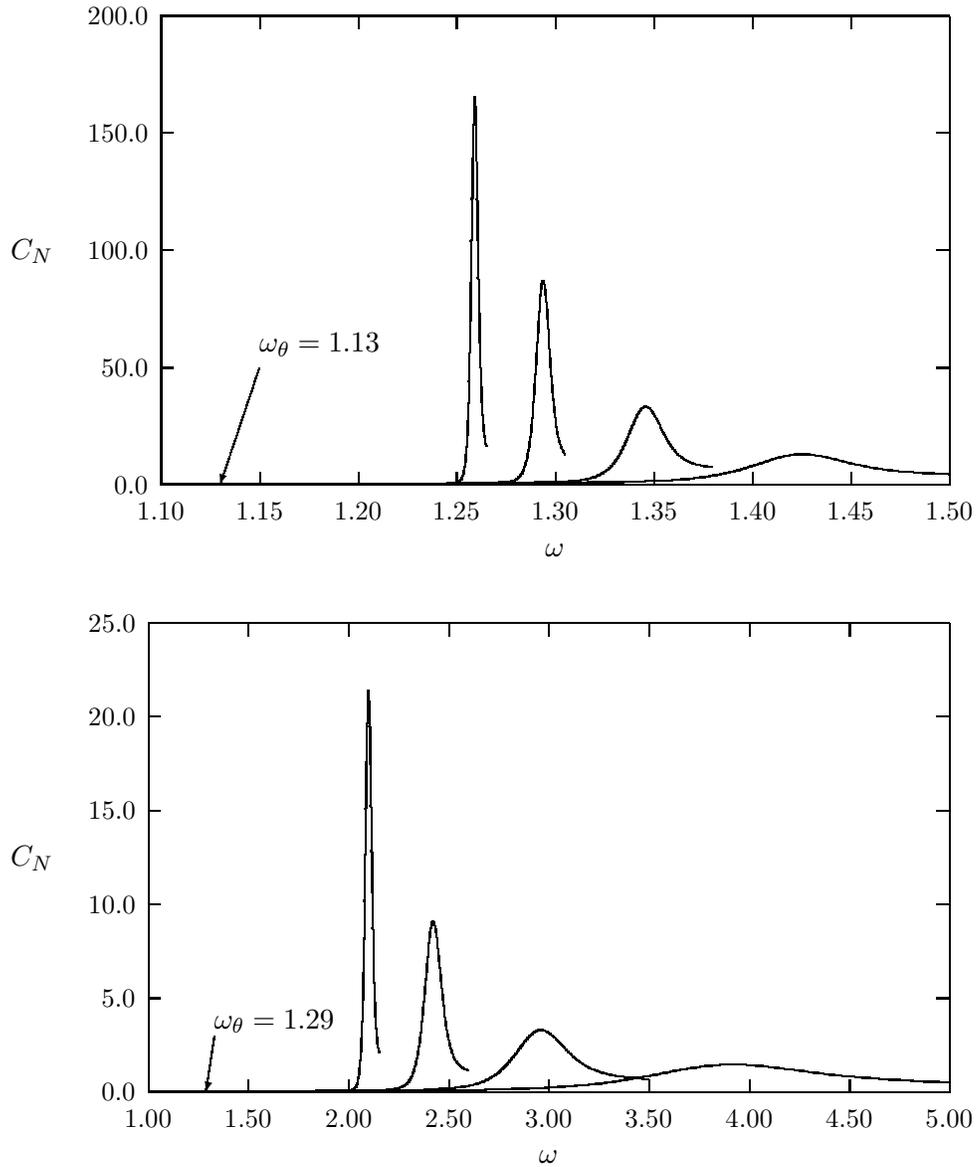

\begin{center}
% GNUPLOT: LaTeX picture
\setlength{\unitlength}{0.240900pt}
\ifx\plotpoint\undefined\newsavebox{\plotpoint}\fi
\sbox{\plotpoint}{\rule[-0.200pt]{0.400pt}{0.400pt}}%
%
\end{center}
\caption{\it Specific heat $C_N$ versus $\omega$ for ISAW at lengths
$256$, $512$, $1024$, and $2048$ (upper figure) and ISAT at lengths $64$,
$128$, $256$, and $512$ (lower figure) from right to left
respectively, using the multi-histogram method.} 
\label{figure3} 
\end{figure}

The location of the collapse transition $\omega_{c,N}$ was found by considering the 
peak of the specific heat curves. For both ISAW and ISAT it is observed that the
specific heat curves display a divergent specific heat peak and a small
transition region of the collapse well separated from the $\theta$-region, see
Figure \ref{figure3}. As we shall see below the region where the specific heat
is large is also the region where the distribution of the internal energy is
bimodal, with different peaks of this distribution dominant at each end of the
temperature range. This indicates that a first-order like transition occurs in
this region.

The collapse region is clearly shifted away from the $\theta$-region at finite
lengths but moves towards the $\theta$-region as length increases. Assuming
there exists a single transition in the thermodynamic limit we now argue that
the collapse point and the $\theta$-point must coalesce as $N$ becomes very
large.  Equation (\ref{shift}) predicts that
$N^{1/4}(\omega_{c,N}-\omega_\theta)$ should approach a constant. Using our
$\theta$-point estimates, this quantity is plotted in Figure \ref{figure4} and
indeed is almost a constant over the range of simulated lengths with a small
linear correction in $N^{-3/4}$. (The chosen scale is due to the fact that we
expect further analytic corrections to scaling in $\omega_{c,N}$.) An
analogous plot using the four-dimensional value for the shift exponent shows
clear curvature. Our data is then consistent with the dimensional dependence
of the shift exponent predicted in (\ref{shift}).

As Equation (\ref{width}) predicts that the width $\Delta\omega$ of the
transition decreases as $N^{-3/4}$, Figure \ref{figure4} also shows the
$N$-dependence of $N^{3/4}\Delta\omega$. While $\Delta\omega$ decreases even
faster than to be expected for even a first-order transition, an asymptotic
scaling of $N^{-3/4}$ as predicted by the PFO theory is not inconsistent with
the data.

As mentioned above, the character of the transition becomes apparent if one
plots the internal energy density distribution (rescaled density of
interactions) at the finite-size collapse transition temperature, $\omega_{c,N}$. Figure
\ref{figure5} shows the emergence of a bimodal distribution for both ISAW and
ISAT.  As lengths increase, the distributions become dominated by two sharp
and well-separated peaks. The values of the minima and maxima of the
distribution are different by three orders of magnitude for the largest
lengths.  As $\omega$ is increased through the transition region the density
distribution switches from the peak located at a small value of contacts to
the peak located at a larger value of contacts, corresponding to a sudden
change in the internal energy.  In the collapsed phase, the width of the peak
is much wider than in the swollen phase, implying a larger specific heat. It
is this difference between the swollen and collapsed phases' specific heats
that will eventually become the thermodynamic second-order jump. The rapid
first-order like switch between two peaks in the distribution becomes more
pronounced at larger polymer lengths since the depth of the ``valley'' between
the two peaks becomes relatively larger.

Continuing with the scaling predictions from the PFO theory, a suitably
defined finite-size latent heat, $\Delta U$, should tend to zero as $N^{-1/4}$
in the thermodynamic limit. One possible measure of this latent heat is given
by the product of specific heat peak $C_N(\omega_{c,N})$ and specific heat peak
width $\Delta\omega$, and another is given by the distance $\delta U$ of the
peaks in the bimodal internal energy distribution.  Figure \ref{figure6} shows
the behaviour of both of these quantities for ISAW and ISAT.  As in four
dimensions, one again notices that even at the longest lengths there is
considerable discrepancy between the two quantities plotted, so that one needs
to be cautious in the interpretation of the scaling behaviour. However the
ISAW data at the largest lengths is beginning to show a consistent decrease in
the latent heat. For ISAT only one measure is so behaved. 
This difficulty is also evident in Figure \ref{figure7}, where the
interfacial tension between the two peaks is shown. For a true first-order
transition this quantity should decrease exponentially in $N$, whereas the PFO
theory predicts an exponential decrease in $N^{1/2}$. The ISAW data is
consistent with an exponential decrease in $N^{1/2}$, and so with the PFO
hypothesis, while there is significant curvature in Figure \ref{figure7} for
the ISAT data. In fact the ISAT data is most consistent with a
decrease in the interfacial tension exponentially in $N$, and so with the
hypothesis of a true first-order transition. We also note that the interfacial
tension is comparatively larger for ISAT than for ISAW, a fact already found
in the four-dimensional simulations \cite{prellberg2001a-:a,prellberg2002a-:a}.
While this results in a stronger transition for ISAT at shorter lengths, the 
first-order nature of this transition makes it more difficult to simulate 
ISAT than ISAW at equal lengths.

In this paper we have discussed the results of large scale Monte Carlo
simulations of interacting self-avoiding walks and trails on the hyper-cubic
lattice in five dimensions. The data was compared to the predictions of a
pseudo first-order transition. The best comparison with the theory came from
the shift of the finite size transition temperature which clearly showed a exponent
difference from previous four-dimensional simulations. The strength of the
transition was found to be even stronger in five than four dimensions. The
difficulty that follows from this implies that further
simulations in even higher dimensions would not be profitable at this
stage. While the ISAW data was reasonably in agreement with the PFO theory for
all the quantities calculated the ISAT data showed a distribution of internal energy that is
consistent with a true first-order transition. We argue though that this is
due to the simulations not reaching the asymptotic regime as demonstrated by
the inconsistency of various measures of the latent heat. To progress further
with present hardware one needs a new algorithm that will both cope with the
first-order-like bimodality of the internal energy distribution and, at the
same time, be an efficient algorithm in the simulation of collapsed
configurations of polymers.

\section*{Acknowledgements}

Financial support from  the Australian Research Council is gratefully
acknowledged by ALO. ALO also thanks the Institut f\"ur Theoretische Physik at the
Technische Universit\"at Clausthal, while TP thanks the  Department of
Mathematics and Statistics at the University of Melbourne: in both cases where
some of the work was completed.

\clearpage

\newpage

\clearpage
\newpage

\clearpage
\newpage

\clearpage
\newpage

\begin{figure}
\begin{center}
% GNUPLOT: LaTeX picture
\setlength{\unitlength}{0.240900pt}
\ifx\plotpoint\undefined\newsavebox{\plotpoint}\fi
\sbox{\plotpoint}{\rule[-0.200pt]{0.400pt}{0.400pt}}%
%
\end{center}
\caption{\it Scaling of the transition: shift and width of the
collapse region. Shown are the scaling combinations 
$N^{1/4}(\omega_{c,N}-\omega_\theta)$ and $N^{3/4}\Delta\omega$ versus
$N^{-3/4}$ for ISAW (upper figure) and ISAT (lower figure).} 
\label{figure4} 
\end{figure}

\clearpage
\newpage

\begin{figure}
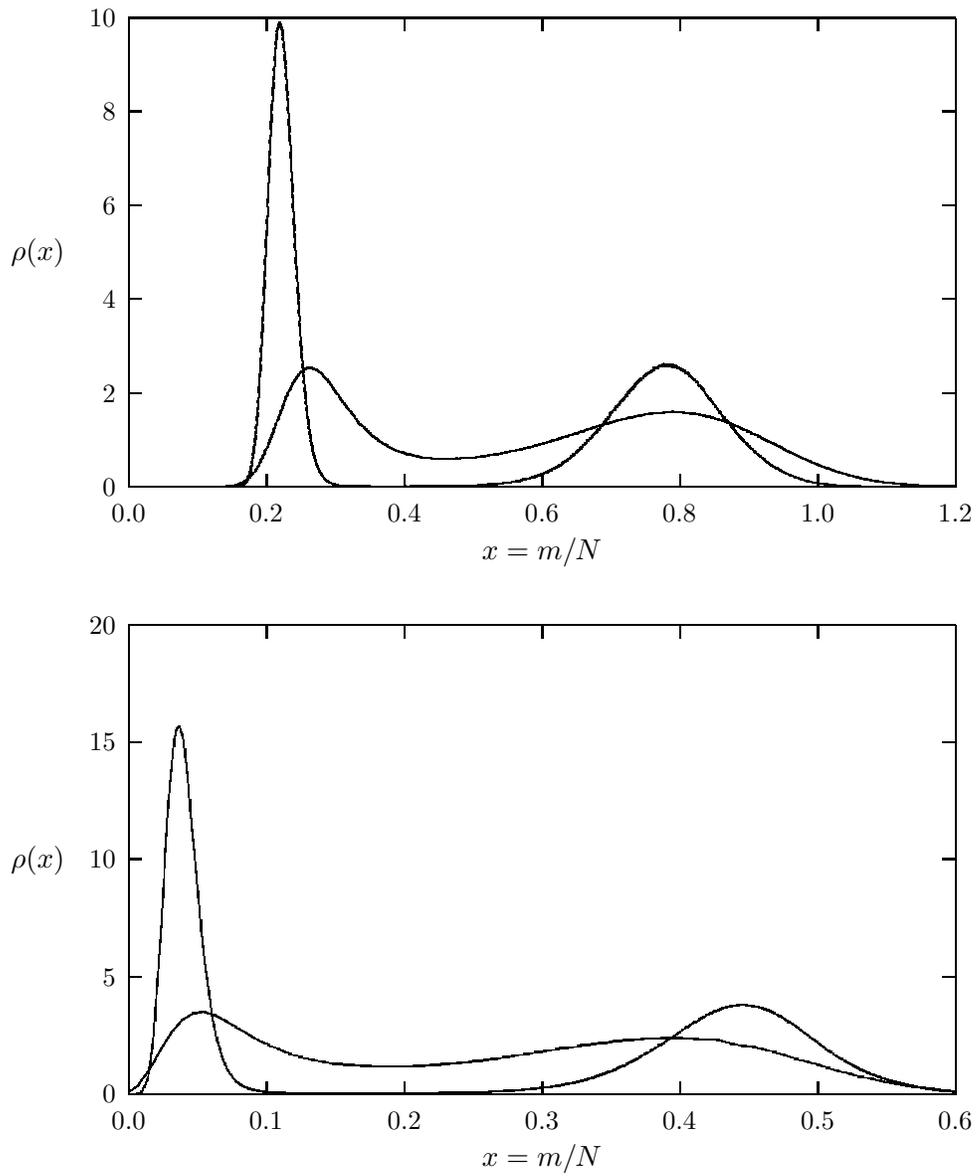

\begin{center}
% GNUPLOT: LaTeX picture
\setlength{\unitlength}{0.240900pt}
\ifx\plotpoint\undefined\newsavebox{\plotpoint}\fi
\sbox{\plotpoint}{\rule[-0.200pt]{0.400pt}{0.400pt}}%
%
\end{center}
\caption{\it Internal energy density distributions at $\omega_{c,N}$.
The upper figure shows data for ISAW at lengths $512$ and $2048$, whereas
the lower figure shows data for ISAT at lengths $128$ and $512$.
The more highly peaked distributions are associated with larger lengths. 
}
\label{figure5} 
\end{figure}

\clearpage
\newpage

\begin{figure}
\begin{center}
% GNUPLOT: LaTeX picture
\setlength{\unitlength}{0.240900pt}
\ifx\plotpoint\undefined\newsavebox{\plotpoint}\fi
\sbox{\plotpoint}{\rule[-0.200pt]{0.400pt}{0.400pt}}%
%
\end{center}
\caption{\it Scaling of the latent heat $\Delta U$ for ISAW (upper figure) and
  ISAT (lower figure):
our two measures of $\Delta U$, $C_N(\omega_{c,N})\Delta\omega$ and
peak distance $\delta U$ are plotted versus $N^{-1/4}$.}
\label{figure6} 
\end{figure}

\clearpage
\newpage

\begin{figure}
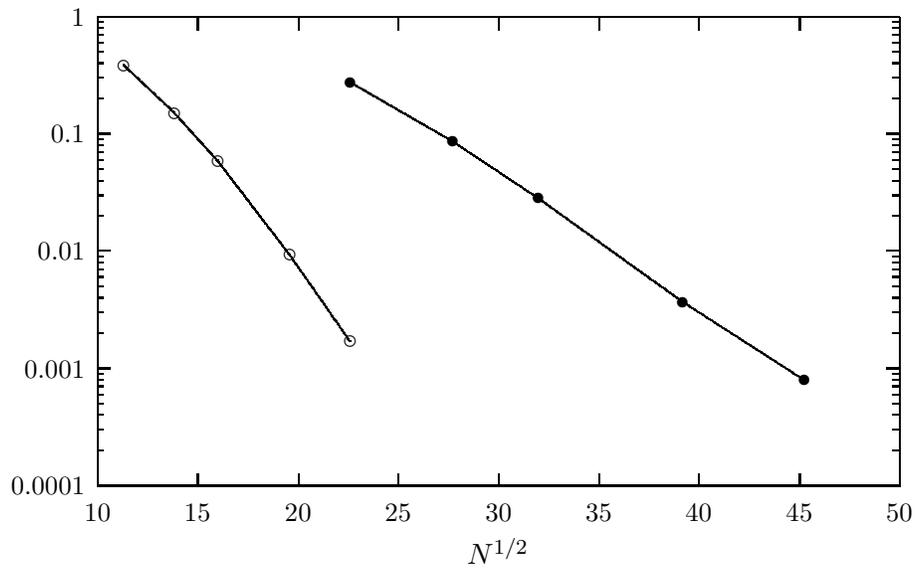

\begin{center}
% GNUPLOT: LaTeX picture
\setlength{\unitlength}{0.240900pt}
\ifx\plotpoint\undefined\newsavebox{\plotpoint}\fi
\sbox{\plotpoint}{\rule[-0.200pt]{0.400pt}{0.400pt}}%
%
\end{center}
\caption{\it ``Interfacial tension'' for ISAW (filled circles) and ISAT
  (empty circles): plotted is the height of the minimum
between the two peaks relative to the height of the maxima (at an
appropriately chosen temperature, at which both peaks are of equal height).
The logarithm of this quantity should scale as $-N(\Delta
U)^2\sim-N^{1/2}$. The ISAW data fall consistently on a straight line, and so
are compatible with this hypothesis, whereas the ISAT data do not. 
}
\label{figure7} 
\end{figure}

\end{document}